\documentclass[preprint]{elsarticle}
\usepackage{graphicx}
\usepackage{amssymb}
\usepackage{amsmath}
\usepackage{natbib}
\usepackage{dcolumn}
\usepackage{bm}
\usepackage{epsfig}
\usepackage{setspace}
\usepackage[caption=false]{subfig}

\journal{Nuclear Physics A}
\begin{document}

\begin{frontmatter}

\title{Formation of droplets with high baryon density at the QCD phase transition in expanding matter}

\author[1,2,3]{Christoph Herold}
\author[3,4]{Marlene Nahrgang}
\author[3,5]{Igor Mishustin}
\author[2,3]{Marcus Bleicher}

\address[1]{School of Physics, Suranaree University of Technology, 111 University Avenue, Nakhon Ratchasima 30000, Thailand}
\address[2]{Institut f\"ur Theoretische Physik, Goethe-Universit\"at, Max-von-Laue-Strasse~1, 
60438 Frankfurt am Main, Germany}
\address[3]{Frankfurt Institute for Advanced Studies (FIAS), Ruth-Moufang-Strasse~1, 60438 Frankfurt am Main, Germany}
\address[4]{Department of Physics, Duke University, Durham, NC 27708, USA}
\address[5]{Kurchatov Institute, National Research Center, 123182 Moscow, Russia}

\begin{abstract}

We consider the (3+1) dimensional expansion and cooling of the chirally-restored and deconfined matter at 
finite net-baryon densities as expected in heavy-ion collisions at moderate energies. 
In our approach, we consider chiral fields and the Polyakov loop as dynamical variables coupled to a medium 
represented by a quark-antiquark fluid. The interaction between the fields and the fluid leads to dissipation 
and noise, which in turn affect the field fluctuations. We demonstrate how inhomogeneities in the net-baryon 
density may form during an evolution through the spinodal region of the first-order phase transition. For 
comparison, the dynamics of transition through the crossover and critical end point is also considered. 

\end{abstract}

\begin{keyword}
 relativistic heavy-ion collisions\sep relativistic fluid dynamics\sep dynamic symmetry breaking\sep fluctuations at first-order phase transitions
\end{keyword}

\end{frontmatter}

\section{Introduction}

Presently it is well established by QCD lattice calculations \cite{Aoki:2006we,Borsanyi:2010bp} that the
chiral/deconfinement phase transition is of the crossover type at small values
of baryon chemical potential $\mu_B$. But a first order phase transition may be
expected at large values of $\mu_B$, starting from a critical end point (CEP).
Such a phase diagram is predicted by several effective models, see e.g.
$\mu_B$ \cite{Scavenius:2000qd,Schaefer:2007pw,Fukushima:2008wg,Herbst:2010rf}, which incorporate some symmetry properties of QCD. It is commonly
believed that the QCD phase transitions can be observed in relativistic
heavy-ion collisions. However, they will be necessarily accompanied by
nonequilibrium phenomena due to a very fast collective expansion of produced
matter. At a first-order phase transition the rapid expansion should lead to
strong supercooling of the deconfined phase. If nucleation times are large, a
fast cooling will drive the system into the unstable region where it may decay via
spinodal decomposition \cite{Mishustin:1998eq,Sasaki:2007qh,Randrup:2010ax,Steinheimer:2012gc}. This can lead to the formation of domains
(droplets) of the decaying phase embedded in the low-density background.

Since the regime of high net-baryon densities is expected at moderate collision energies, it 
will be accessible in future experiments at FAIR 
\cite{Friman:2011zz} and NICA \cite{nica:whitepaper}. There we will have the opportunity to directly 
investigate the first-order phase transition by looking for potential signals
stemming from the spinodal dynamics.

In this article we present a fully dynamical approach to the formation of domains with high 
baryon density at the first-order 
phase transition during a heavy-ion collision. For this purpose we apply the recently developed model of 
nonequilibrium chiral fluid dynamics \cite{Nahrgang:2011mg,Nahrgang:2011mv,Nahrgang:2011vn,Herold:2013bi}.  
It is based on earlier studies of chiral fluid dynamics, where the chiral order parameter was explicitly 
propagated via the classical Euler-Lagrange equation \cite{Mishustin:1998yc,Paech:2003fe}. 
A locally thermalized background medium was given by a fluid of quarks and antiquarks, taking into account 
energy and momentum exchange between field and fluid. 

The nonequilibrium version of the model includes damping processes due to 
the interaction of the field with the fluid and the back-reaction via stochastic noise, incorporated in a Langevin equation
\cite{Nahrgang:2011mg}. 
This enables us to describe relaxational processes which do not appear in the classical Euler-Lagrange 
dynamics. Using this formalism the equilibration processes in a finite system were studied in \cite{Nahrgang:2011mv}, and numerical results 
of the fully (3+1) dimensional fluid dynamic expansion were presented in \cite{Nahrgang:2011vn}. They include a 
strong enhancement of nonequilibrium fluctuations at the first-order phase transition as well as a reheating of the 
quark fluid after the decay of a supercooled phase.

In \cite{Herold:2013bi} we extended the chiral fluid dynamics model by including an explicit
propagation of an effective Polyakov loop field. This enables us to describe nonequilibrium
effects in combined dynamics of chiral and deconfinement order parameters. We
have demonstrated that soft modes of the sigma field are indeed strongly
enhanced at the critical point, if the system is equilibrated in a box. It is, however, well 
known that due to critical slowing down critical fluctuations do not fully develop in a 
dynamical situation. This was phenomenologically studied in \cite{Berdnikov:1999ph}. Applied to a system expanding and cooling through a first order
phase transition, the formation of domains with high and low values of order
parameters was clearly observed.

In the present work we extend these studies to the case of finite baryochemical
potential to study the domain formation in systems with finite baryon
densities. The dynamical trajectories of the fluid elements are now
two-dimensional in the (T-$\mu$)-plane. This has a significant effect on
the dynamics at intermediate times when the system is trapped in the spinodal
region for a noticeable time. Furthermore, now we have to deal with a realistic baryon-rich fluid,
characterized by the energy density and the net-baryon density. While
the first quantity may fluctuate due to a stochastic source term, the latter one
is only subject to a continuity equation. It will be interesting to see, whether
the formation of non-uniform structures that we observed in \cite{Herold:2013bi} for
the energy density will also be possible in the baryon density, which is
connected to the baryon number distribution.

After an introduction to the model we derive the equations of motion in Sec.~\ref{sec:model}. 
Sec.~\ref{sec:trajectories} presents nonequilibrium trajectories for different 
initial conditions in the ($T$-$\mu$)-plane. In Sec.~\ref{sec:bubbles}, we show how nonequilibrium effects at the 
first-order phase transition lead to clustering in the baryon density, which can be observed by non-statistical 
azimuthal fluctuations of the baryon density. The paper concludes with a summary and 
an outlook in Sec.~\ref{sec:summary}.

\section{Nonequilibrium chiral fluid dynamics}
\label{sec:model}

The model is built on the Lagrangian of the Polyakov-Quark-Meson (PQM) model \cite{Schaefer:2007pw}
\begin{equation}
\label{eq:Lagrangian}
\begin{split}
{\cal L}&=\overline{q}\left[i \left(\gamma^\mu \partial_\mu-i g_{\rm s}\gamma^0 A_0\right)-g \left(\sigma +i\gamma_5 \vec\tau\cdot \vec\pi\right)\right]q \\
&\hphantom{=} + \frac{1}{2}\left(\partial_\mu\sigma\right)^2 + \frac{1}{2}\left(\partial_\mu\vec\pi\right)^2 
- U\left(\sigma,\vec\pi\right) - {\cal U}_T(\ell, \bar\ell)~.
\end{split}
\end{equation}
Similar to the Polyakov loop extended Nambu-Jona-Lasinio model 
\cite{Fukushima:2003fw}, it combines aspects of confinement 
with a model for chiral symmetry breaking by the inclusion of a static homogeneous temporal color 
gauge field $A_0$. From both models, conclusions have been drawn on the existence of a first-order phase transition at large baryochemical potential. Within the chiral fluid dynamics approach, we offer a dynamical version of these models. 
In the PQM model the mesonic fields $\sigma$ and $\vec{\pi}$ couple to the quarks and antiquarks with a 
coupling strength $g$. The temporal component of the color gauge field also couples to the quarks, generating 
the Polyakov loop $\ell$ in the mean-field effective potential $V_{\rm eff}$ which is represented as
\begin{equation}
 V_{\rm eff}=U(\sigma)+{\cal U}_T(\ell,\bar\ell)+\Omega_{\rm q\bar q}(\sigma,\ell,\bar\ell,T,\mu)~.
\end{equation}
It is obtained from the grand canonical partition function by integrating out the quark degrees of freedom 
which will constitute the heat bath. 
Here, $U$ is the classical Mexican hat potential of the mesonic fields
\begin{equation}
U\left(\sigma\right)=\frac{\lambda^2}{4}\left(\sigma^2-\nu^2\right)^2-h_{\rm q}\sigma-U_0~.
\label{eq:Usigma}
\end{equation}
Chiral symmetry is spontaneously broken in the vacuum, where $\langle\sigma\rangle=f_\pi=93$~MeV, the pion decay constant. 
The explicit symmetry breaking term is set to $h_q=f_\pi m_\pi^2$ with the pion mass $m_\pi=138$~MeV. With this we have 
$\nu^2=f_\pi^2-m_\pi^2/\lambda^2$. 
We choose $\lambda^2=19.7$ to obtain a realistic sigma mass of $m_\sigma^2=2\lambda^2 f_\pi^2 + m_\pi^2\approx (600\text{ MeV})^2$ in the vacuum. 
The constant term $U_0=m_\pi^4/(4\lambda^2)-f_\pi^2 m_\pi^2$ shifts the potential energy such that it vanishes in the 
ground state. The quark-meson coupling constant $g$ is set to a value of $3.3$ to reproduce the vacuum constituent quark 
mass.

A temperature-dependent effective potential of the Polyakov loop is defined as 
\begin{equation}
\frac{{\cal U}_T}{T^4}\left(\ell, \bar\ell\right)= -\frac{b_2(T)}{4}\left(\left|\ell\right|^2+\left|\bar\ell\right|^2\right)-\frac{b_3}{6}\left(\ell^3+\bar\ell^3\right) + \frac{b_4}{16}\left(\left|\ell\right|^2+\left|\bar\ell\right|^2\right)^2~.
\label{eq:Uloop}
\end{equation}
It is obtained from fits of the pressure to lattice QCD data in the pure Yang-Mills sector 
\cite{Pisarski:2000eq} which leads to a temperature dependent 
coefficient $b_2$
\begin{equation}
 b_2(T) = a_0 + a_1\left(\frac{T_0}{T} \right) + a_2\left(\frac{T_0}{T} \right)^2 + a_3\left(\frac{T_0}{T} \right)^3
\end{equation}
and the choice of parameters $a_0=6.75$, $a_1=-1.95$, $a_2=2.625$, $a_3=-7.44$ together with $b_3=0.75$ and 
$b_4=7.5$.

Finally, the mean-field thermodynamic potential of the quarks reads
\begin{equation} 
  \begin{split}
\label{eq:grandcanonical}
\Omega_{\rm q\bar q}
=&-2 N_f T\int\frac{\mathrm d^3 p}{(2\pi)^3} \\
&\left\{\ln\left[g^{(+)}(\sigma,\ell,T,\mu)  \right]+\ln\left[g^{(-)}(\sigma,\ell,T,\mu)  \right]\right\}~.
  \end{split}
\end{equation}
Here we used the following notation
\begin{eqnarray}
g^{(+)}&=&1+3\ell\mathrm e^{-E^{(+)}/T}+3\ell\mathrm e^{-2E^{(+)}/T}+\mathrm e^{-3E^{(+)}/T}, \\
g^{(-)}&=&1+3\ell\mathrm e^{-E^{(-)}/T}+3\ell\mathrm e^{-2E^{(-)}/T}+\mathrm e^{-3E^{(-)}/T},
\end{eqnarray}
with the number of flavors $N_f=2$ and $E^{(+)}=E_{\rm q}-\mu$, $E^{(-)}=E_{\rm q}+\mu$, where $\mu=\mu_B /3$ 
is the quark chemical potential . Throughout this work, we neglect pion fluctuations and concentrate on the 
sigma field and the Polyakov loop, which are the order parameters for the chiral and deconfinement 
transitions. The energy of the quarks $E_{\rm q}=\sqrt{\vec p^2+m_{\rm q}^2}$ depends on the sigma field via the 
dynamically generated effective mass $m_{\rm q}=g|\sigma|$. In the mean-field approximation, the chiral and 
deconfinement transition temperatures coincide for all values of $\mu$. 
We also find a common CEP at $(T_{\rm cp},\mu_{\rm cp})=(152,160)$~MeV. 

For effective models of the present type it was found in \cite{Csernai:1992tj,Palhares:2010be} that nucleation times are 
much larger than typical expansion times of the fireball, which implies that the system will strongly 
supercool. This conclusion was confirmed within the present model in \cite{Nahrgang:2011mv,Herold:2013bi}, 
thus giving us the ground to expect significant effects stemming from spinodal decomposition. 

We now briefly describe the coupled dynamics of the fields and the fluid.
In our approach the sigma field is propagated according to a Langevin equation
\begin{equation}
\label{eq:eomsigma}
 \partial_\mu\partial^\mu\sigma+\eta_{\sigma}\partial_t \sigma+\frac{\delta V_{\rm eff}}{\delta\sigma}=\xi_{\sigma}~,
\end{equation}
which besides the standard mean-field contribution contains the damping term with friction coefficient $\eta_{\sigma}$ and the 
stochastic noise field $\xi_{\sigma}$. For $m_\sigma(T)>2m_q(T)=2g\sigma_{\rm eq}(T)$ the damping coefficient 
due to the decay of the zero mode sigma field into a quark-antiquark pair ($\sigma\to q+\bar q$) is given by 
\cite{Nahrgang:2011mg} 
\begin{equation}
\label{eq:dampingcoeff}
  \eta_{\sigma}=\frac{12 g^2}{\pi}\left[1-2n_{\rm F}\left(\frac{m_\sigma}{2}\right)\right]\frac{1}{m_\sigma^2}\left(\frac{m_\sigma^2}{4}-m_q^2\right)^{3/2}~,
\end{equation}
where $n_{\rm F}(k)$ is the Fermi occupation number. 
Since we are mainly interested in the long-range fluctuations of the sigma field, we use the damping coefficient 
(\ref{eq:dampingcoeff}) for all modes in the Langevin equation. At lower temperatures it vanishes because 
the decay into a quark-antiquark pair is kinematically forbidden as $m_\sigma(T)<2m_q(T)=2g\sigma_{\rm eq}(T)$. However, the sigma 
field is further damped due to interaction with the hard chiral modes \cite{Greiner:1996dx,Rischke:1998qy}, 
which can be represented by the reaction $\sigma\leftrightarrow2\pi$. To account for them we use the value 
$\eta=2.2/{\rm fm}$ \cite{Biro:1997va} for the kinematic range $2m_q>m_\sigma(T)>2m_\pi$.

The stochastic field in the Langevin equation (\ref{eq:eomsigma}) is given by a Gaussian distribution with 
zero mean $\langle\xi_{\sigma}(t)\rangle_\xi=0$.
The amplitude of the noise is determined by the dissipation-fluctuation theorem
\begin{equation}
\label{eq:dissfluctsigma}
 \langle\xi_{\sigma}(t,\vec x)\xi_{\sigma}(t',\vec x')\rangle_\xi=\delta(\vec x-\vec x')\delta(t-t')m_\sigma\eta_{\sigma}\coth\left(\frac{m_\sigma}{2T}\right)~.
\end{equation}
The friction and noise terms ensure relaxation of the order parameter to the proper equilibrium distribution in the stable phase region. 

\begin{figure}[t]
\centering
    \includegraphics[scale=0.65]{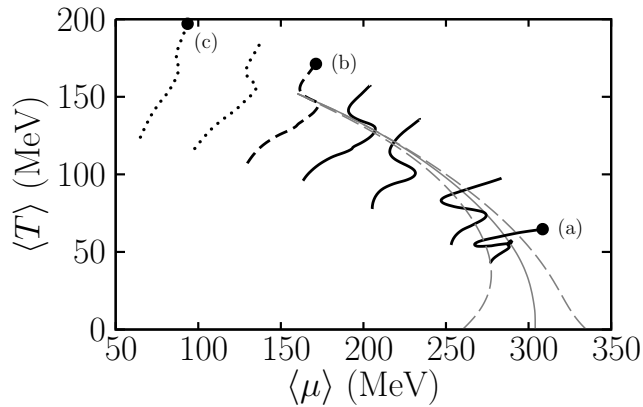}
\caption{Phase diagram of the model in ($T$-$\mu$)-plane: thin solid line shows the boundary of
the first order phase transition, while dashed lines delineate the spinodal
region. Various bold lines show dynamical trajectories of the quark fluid in the
($T$-$\mu$)-plane for different initial conditions. The averaging is made over the central box of $1\,\rm{fm}^3$.}
\label{fig:trajectories}
\end{figure}

The dynamics of the Polyakov loop $\ell$ is described phenomenologically by the relaxation equation
\begin{equation}
\label{eq:eomell}
 \eta_{\ell}T\partial_t \ell +\frac{\delta V_{\rm eff}}{T\delta\ell}=\xi_{\ell}~,
\end{equation}
with a constant damping coefficient $\eta_{\ell}=5/{\rm fm}$. This value can only be estimated very roughly, 
but the results are largely independent of its specific choice \cite{Herold:2013bi}. A rigorous derivation 
of the equation of motion is not possible, as $\ell$ is originally defined in Euclidean space 
with imaginary time axis used for the temperature. The often used kinetic term $\sim T^2\vert\partial_\mu \ell\vert^2$ \cite{Dumitru:2000in,Dumitru:2001bf} 
is problematic for the current studies as the temperature field is non-uniform and time-dependent, creating 
derivative terms of $T$ in the Euler-Lagrange equation of motion which may lead to unphysical behavior. 

The stochastic noise 
$\xi_{\ell}$ is assumed to be Gaussian, too, and we impose the dissipation-fluctuation relation \cite{Herold:2013bi}
\begin{equation}
\label{eq:dissfluctpolyakov}
 \langle\xi_{\ell}(t,\vec x)\xi_{\ell}(t',\vec x')\rangle_\xi=\delta(t-t')\delta(\vec x-\vec x')2\eta_{\ell} T~.
\end{equation}

In our dynamical model, the expansion of the fireball is described according to energy-momentum and baryon 
number conservation equations
\begin{eqnarray}
\label{eq:fluidT}
\partial_\mu T_{\rm q}^{\mu\nu}(t,\vec x)&=&S^\nu(t,\vec x)~,\\
\label{eq:fluidN}
\partial_\mu N^{\mu}(t,\vec x)&=&0~,
\end{eqnarray}
where $T_{\rm q}$, the energy-momentum tensor of the quarks, acquires ideal fluid form and 
the source term $S^\nu=-\partial_\mu\left(T_\sigma^{\mu\nu}+T_\ell^{\mu\nu}\right)$ accounts for the energy-momentum exchange between the fluid and the fields.
It is dominated by the energy dissipation from the sigma field and Polyakov loop to the fluid, controlled by 
the damping terms in Eqs.\ (\ref{eq:eomsigma}) and (\ref{eq:eomell}). This energy transfer is largest during 
the relaxation from the supercooled state. For the baryon number current, we have $N^{\mu}=n u^\mu$, 
with the local baryon number density $n$ and the four-velocity of the fluid $u^{\mu}=\gamma(1,\vec v)$.

Since the fields $\sigma$ and $\ell$ are not necessarily at their respective equilibrium values, the equation 
of state will depend explicitly on their local values $\sigma(\vec x)$, $\ell(\vec x)$. This is different from 
standard fluid dynamical simulations using equilibrium equations of state. The corresponding equation of state 
of the fluid is obtained from the generalized thermodynamic relations
\begin{align}
\label{eq:pressure}
 p(\sigma,\ell,T,\mu)&=-\Omega_{\rm q\bar q}~,\\
\label{eq:energydensity}
 e(\sigma,\ell,T,\mu)&=T\frac{\partial p}{\partial T}
 + \mu\frac{\partial p}{\partial \mu}-p~,\\
\label{eq:quarkdensity}
 n(\sigma,\ell,T,\mu)&=\frac{\partial p}{\partial\mu}~.
\end{align}
where $\sigma$ and $\ell$ are functions of time and coordinates due to their own dynamics.

\section{Trajectories in the phase diagram}
\label{sec:trajectories}

We initialize a spherical droplet of quark matter in the center of a numerical grid with $128^3$ cells. We choose different values of the 
initial temperature $T_{\rm ini}$ and quark chemical potential $\mu_{\rm ini}$ and use a Woods-Saxon function to 
ensure a smooth transition to the vacuum at the edges of the droplet 
\begin{equation}
 T(t=0,\vec x)=\frac{T_{\rm ini}}{1+\exp[(\left|\vec x\right|-R)/ a]}~,~~
 \mu(t=0,\vec x)=\frac{\mu_{\rm ini}}{1+\exp[(\left|\vec x\right|-R)/ a]}~.
\end{equation}
Here $R=4.0$~fm denotes the radius of the sphere and $a=0.5$~fm the surface thickness. The fields $\sigma$ and $\ell$ are 
initialized according to 
their respective $T$- and $\mu$-dependent equilibrium values with Gaussian fluctuations around them. Finally, 
the thermodynamic quantities $p$, $e$ and 
$n$ are calculated using equations (\ref{eq:pressure}), (\ref{eq:energydensity}) and (\ref{eq:quarkdensity}).

The fluid is propagated numerically using the (3+1)d SHarp And Smooth Transport Algorithm (SHASTA) ideal fluid dynamic 
code \cite{Rischke:1995ir,Rischke:1995mt} with a time step of $\Delta t = 0.4\cdot\Delta x = 0.08$~fm to ensure 
numerical stability. The Langevin equations of motion for the sigma field and Polyakov loop are solved with a staggered 
leap-frog algorithm as it is presented in \cite{CassolSeewald:2007ru}. For this the time step has to be chosen smaller  
than for the fluid, we take that value as
$\Delta t=0.1\cdot\Delta x = 0.02$~fm. That means we perform four consecutive steps of propagating the fields followed 
by one step for the fluid. After that we calculate the local temperature and chemical potential $T(\vec x)$, $\mu(\vec x)$ 
via a root finder of 
\begin{align}
\label{eq:eroot}
 e_{\rm fluid}(\vec x)-e\{\sigma(\vec x), \ell(\vec x), T(\vec x), \mu(\vec x)\} &= 0~,\\
\label{eq:nroot}
 n_{\rm fluid}(\vec x)-n\{\sigma(\vec x), \ell(\vec x), T(\vec x), \mu(\vec x)\} &= 0~.
\end{align}
These temperatures and chemical potentials are then used in the next time step for the propagation of the order parameter fields according 
to equations (\ref{eq:eomsigma}) and (\ref{eq:eomell}).

The noise fields are sampled for each new time step according to the Gaussian distribution, given by Eqs.\ 
(\ref{eq:dissfluctsigma}) and (\ref{eq:dissfluctpolyakov}). For each grid cell ($i\in 1\dots N$) we average the 
initially uncorrelated noise variables over a surrounding volume $V^{\rm corr}_{\sigma}=1/m_{\sigma}^3$ and 
$V^{\rm corr}_{\ell}=1/m_{\ell}^3$ and attribute this average value to the noise in the respective grid cell $i$. 
Thus, the noise values in neighboring cells $i$ and \mbox{$i+1$}, which share large parts of the averaging volume, are 
correlated. This averaging reduces the variance of the noise and we need to recover the original local variance by 
rescaling. In this way we avoid an artificial correlation length equal to the cell spacing. 
In \cite{Herold:2013bi} we have found that during the simulation, the correlation length reaches values of $1.5-2.0$~fm 
around the CEP and remains small, at about $0.2-0.4$~fm, in the case of a first-order transition.

We run the simulation for several initial conditions $(T_{\rm ini},\mu_{\rm ini})$ to probe the crossover, 
CEP and first-order phase transition regime. 

The trajectories of the dynamic simulations in the ($T$-$\mu$)-plane are presented in Fig.~\ref{fig:trajectories} together with the 
corresponding phase boundary and the spinodal lines. The averages $\langle T\rangle$ and $\langle\mu\rangle$ are calculated 
over a central box of $1\,\rm{fm}^3$. The bending of the curves towards larger values of $\mu$ is due to the 
rapid increase in the quark mass at the chiral transition. Similar behavior is already inherent in the equilibrium 
isentropes, which follow the phase boundary towards larger $\mu$ \cite{Scavenius:2000qd,Kahara:2008yg}. 
In our dynamical simulations we see that at larger chemical 
potentials the bending only occurs after the trajectories have crossed the lower spinodal line. 
This behavior has a simple explanation: After crossing the first-order phase transition line, the system is still not able to overcome the potential 
barrier but gets trapped in a metastable chirally restored and deconfined state. After that barrier has vanished at the lower spinodal line, 
the system can roll down to the global minimum where the constituent quarks acquire their mass. 
This process develops in the time interval between two
turning points of the trajectory. At the right turning point the system almost
reaches the equilibrium phase boundary (thin solid line). Further on the
trajectory follows the stable minimum of the effective potential corresponding
to the chirally-broken phase. One should bear in mind that these trajectories
are obtained by averaging over many individual histories and therefore do not
reveal strong fluctuations. At highest $\mu$, trajectory (a) in Fig. 1, one can
even observe a slight reheating of the fluid, an effect which we have already
discussed in refs. \cite{Nahrgang:2011vn,Herold:2013bi}. In summary we see that due to 
nonequilibrium effects the system supercools and spends an extended amount of time in the metastable state. This may 
facilitate the process of spinodal decomposition and therefore the development of interesting effects of the 
first-order phase transition.

\section{Formation of high-density droplets}
\label{sec:bubbles}

\begin{figure}[t]
\centering
    \subfloat[\label{fig:relden4_t6}]{
    \centering
    \includegraphics[scale=1.2]{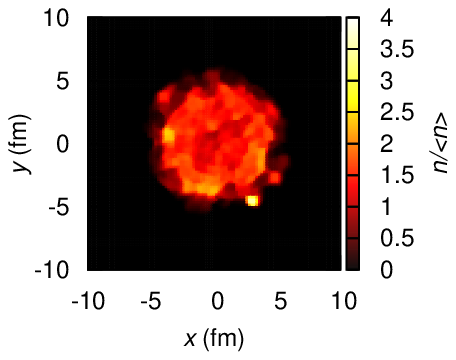}
    }
  \hfill
    \subfloat[\label{fig:relden4_t12}]{
    \centering
    \includegraphics[scale=1.2]{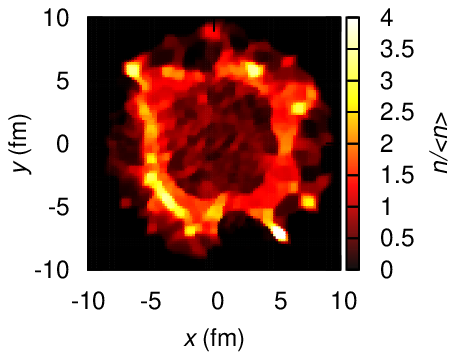}
    }
  \vfill
    \subfloat[\label{fig:relden125_t6}]{
    \centering
    \includegraphics[scale=1.2]{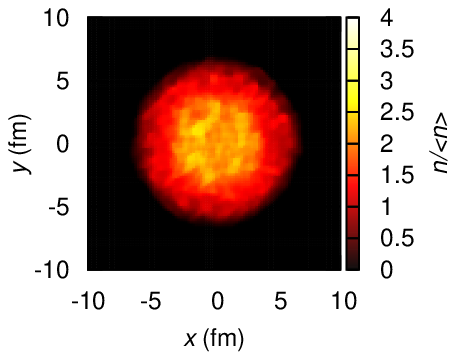}
    }
  \hfill
    \subfloat[\label{fig:relden125_t12}]{
    \centering
    \includegraphics[scale=1.2]{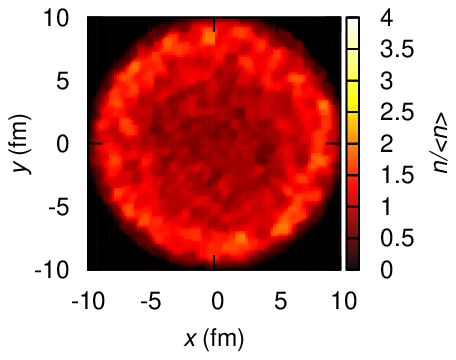}
    }
\caption[Relative net-baryon number]{Relative net-baryon number density at $z=0$ for a first-order transition at 
$t=6$~fm \subref{fig:relden4_t6} and $t=12$~fm \subref{fig:relden4_t12} and for a transition through the CEP for 
$t=6$~fm \subref{fig:relden125_t6} and $t=12$~fm \subref{fig:relden125_t12}. One can see that droplets of high 
density are formed at the first-order phase transition, while the density fluctuations remain small during 
an evolution through the CEP.}
\label{fig:evol4}
\end{figure}

We performed simulations for the expansion of the fireball in the two scenarios with a first-order phase 
transition and with a CEP. They correspond to the trajectories shown by the most right solid line (a) 
and the dashed line (b) in Fig.~\ref{fig:trajectories}. 
In Fig.~\ref{fig:evol4}, we show two-dimensional maps (through the center of the fireball) of the 
relative net-baryon number density $n/\langle n\rangle$, where $\langle n\rangle$ is 
the volume average over all cells with $n>0$. 
We see that during the expansion the fluid evolves non-uniformly in the case of a first-order transition, 
creating droplets of high baryon density at the periphery of the system. 
This effect is not observed at the CEP, where the spherical symmetry is preserved and no strong fluctuations 
occur. It is remarkable that such droplets form in the net-baron density, which is in contrast to the 
energy density not coupled to an external stochastic source term. We conclude that the fluctuations 
in the order parameters induce fluctuations in the fluid dynamical quantities via the equation of state, 
even without any explicit coupling.

\begin{figure}[t]
\centering
    \subfloat[\label{fig:angrho6}]{
    \centering
    \includegraphics[scale=0.5]{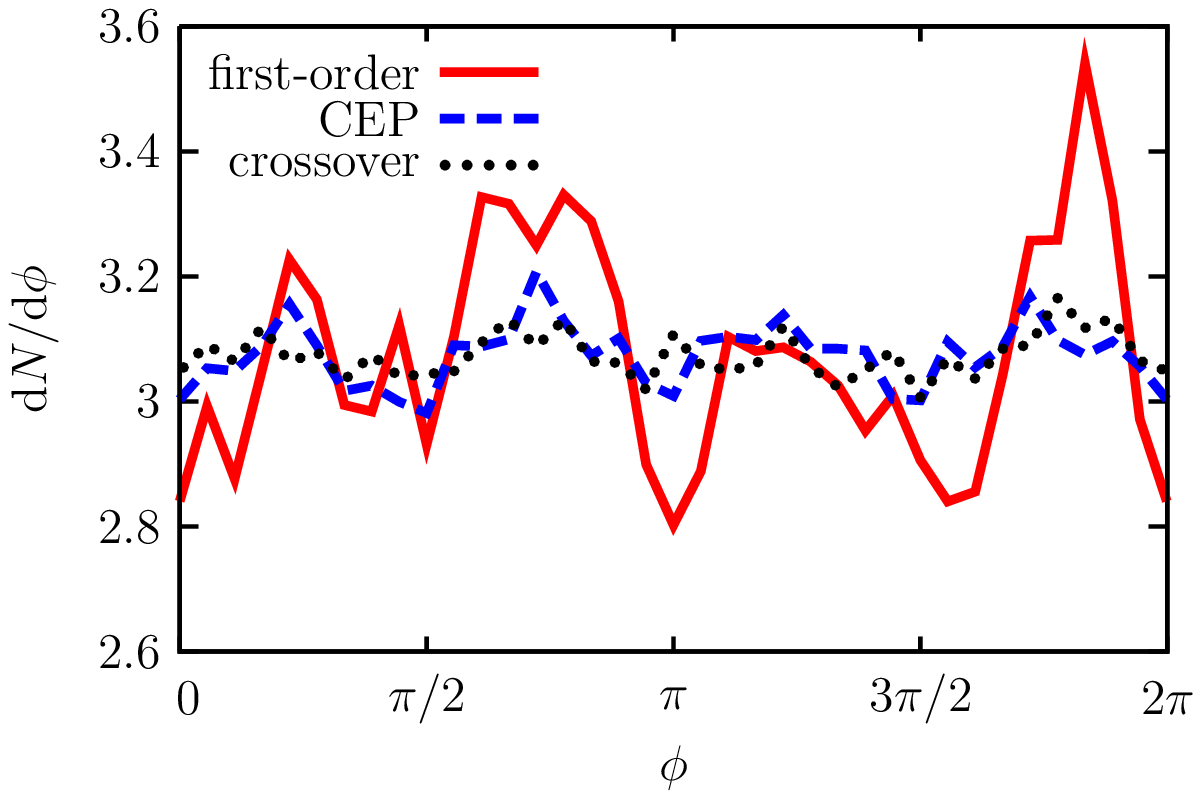}
    }
  \hfill
    \subfloat[\label{fig:angrho12}]{
    \centering
    \includegraphics[scale=0.5]{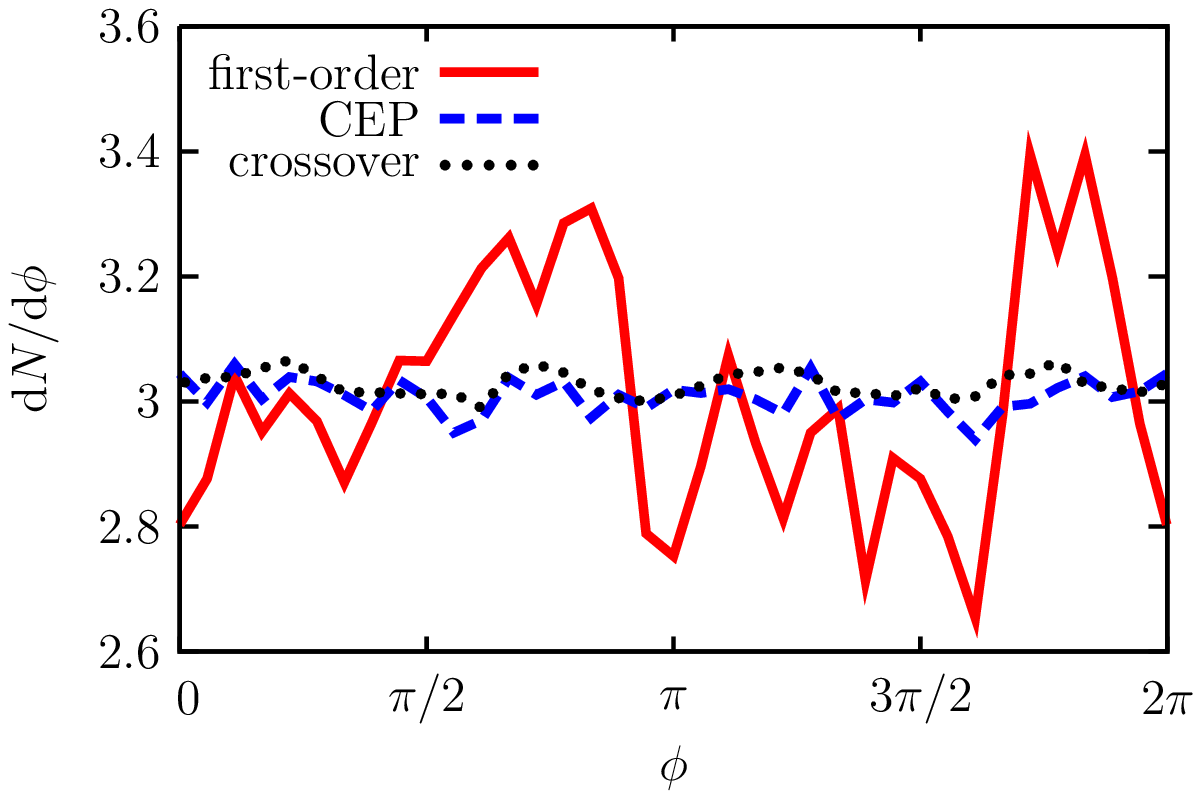}
    }
\caption[Azimuthal distribution of the baryon number]{Azimuthal distribution of the baryon number density after $t=6$~fm \subref{fig:angrho6} and 
$t=12$~fm \subref{fig:angrho12} for several transition scenarios. We see that strong inhomogeneities develop at the 
first-order phase transition.}
\label{fig:angrho}
\end{figure}

It is possible that such inhomogeneities lead to large fluctuations of observables in individual events. 
From Fig.~\ref{fig:evol4} one can see that individual droplets move radially outwards while their relative strength is 
amplified, i.\ e.\ their density grows in comparison with the average density in the volume. This is due to the spinodal instabilities, allowing inhomogeneities to grow during the 
evolution. From these figures we expect that after freeze-out the particles stemming from the 
droplets will show a characteristic bumpy azimuthal distribution 
of the net-baryon number $\mathrm d N/\mathrm d\phi$, where $\phi$ is the azimuthal angle around the 
$z$-axis. Although we have no freeze-out in our model, we have by Eq.\ (\ref{eq:fluidN}) a conserved baryon number which  is not altered by a hadronization process. Such angular distributions are shown in Fig.~\ref{fig:angrho} for $t=6$~fm and $t=12$~fm, corresponding to 
initial conditions (a) for a first-order phase transition, (b) for the CEP and (c) for a crossover from 
Fig.~\ref{fig:trajectories}. One can clearly see 
strong fluctuations of the net-baryon number as a function of $\phi$ at the first-order phase transition in contrast 
to the CEP and crossover. One can notice the correlation of bumps and deeps between the two plots in 
Fig.~\ref{fig:angrho}, which indicate that the droplets preserve their identity during 
this time interval. It is at this point important to note that the stability of these quark droplets
is a direct consequence of the equation of state obtained from the PQM model. As
well known, see e.g. \cite{Scavenius:2000qd,Steinheimer:2013xxa}, at low temperatures this model leads to negative
pressures and the pressure difference between quark matter and the vacuum tends to zero. This means that the quark 
droplets can coexist with the vacuum at zero pressure. For a more realistic equation of state which has no zero-pressure
point, inhomogeneities in the quark density will exist only in the mixed phase
and dissolve after some time. But if the hadronization transition is close to
the freeze-out stage, the strong correlations in the net-baryon distributions
will survive in observables, see discussion in \cite{Mishustin:2007jx}.

\begin{figure}[t]
\centering
    \includegraphics[scale=0.65,angle=270]{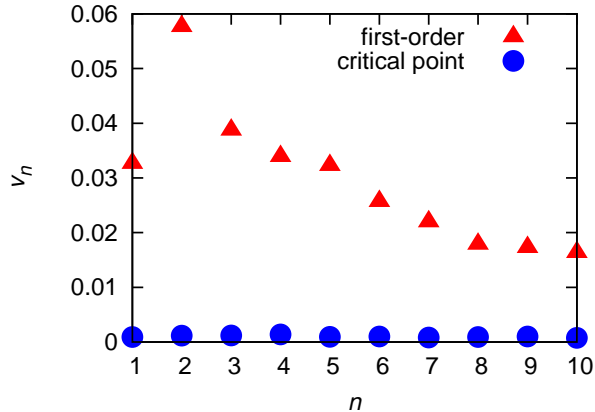}
\caption{Event-averaged Fourier coefficients of net-baryon distributions in position space at $t=12$~fm. Results are shown for the two scenarios: 
evolution through the first-order phase transition (triangles) and through the CEP (dots).}
\label{fig:flow}
\end{figure}

In Fig.~\ref{fig:flow} we show the impact of the formation of droplets on baryonic flow. For this purpose we 
calculated the Fourier coefficients $v_n$ of the angular distributions of net-baryons in position space, $\mathrm d N/\mathrm d\phi$, 
at end of the evolution for $t=12$~fm. They resemble the flow harmonics of the net-baryon distribution measured in experiments. 
To suppress statistical fluctuations these coefficients are averaged over an ensemble of generated events. The results 
are presented in Fig.~\ref{fig:flow}. As one may already expect from the previous discussions, these coefficients are significantly 
larger after an evolution through the first-order phase transition than through the CEP. The largest enhancement is found for $n=2$ harmonic. 
This corresponds to the plot of the azimuthal distribution in Fig.~\ref{fig:angrho}, which shows two large peaks around $\phi=3\pi/4$ and 
$\phi=7\pi/4$ for both the intermediate and the final stage of the simulation. The higher harmonics ($n>2$) are also significantly enhanced 
in the first-order transition scenario.

\section{Summary and Outlook}
\label{sec:summary}

In conclusion, within a fully dynamical approach we have demonstrated how strong
inhomogeneities of baryon density can be created at the first-order QCD
phase transition. They are not caused by the initial fluctuations but appear as
the result of the stochastic evolution of the system during spinodal
decomposition. Within the setup of nonequilibrium chiral fluid dynamics we have
shown that non-uniform structures (droplets) can develop and survive at the
first-order phase transition during the expansion stage. These non-statistical
fluctuations may be observed in net-baryon azimuthal distributions in single
events. In a rapidly expanding system the fluctuations in coordinate space
should translate into fluctuations in momentum space which may survive even
after hadronization of the droplets. Then a prominent signature would be an 
enhancement of higher harmonics of the net-baryon flow, as illustrated in 
Fig.~\ref{fig:flow}. By varying the beam
energy one can get closer to the CEP where the signals of the first-order phase
transition become weaker, until they finally cease.

In the future we are going to extend this work by studying the influence of
initial fluctuations which are present in the initial stage of a heavy-ion
collision. It will be interesting to investigate their evolution in the presence
of spinodal decomposition and their significance in the presence of the
dynamically generated stochastic fluctuations. Moreover, we aim at improving 
the underlying equation of state by a more realistic description of the hadronic phase.

\section*{Acknowledgements}

The authors thank Stefan Leupold, Jan Steinheimer, Stefan Schramm and Carsten Greiner for fruitful discussions 
and Dirk Rischke for providing the SHASTA code. 
This work was supported by the Hessian LOEWE initiative Helmholtz International Center for FAIR. I.~M. 
acknowledges partial support from grant NSH-215.2012~2 (Russia).

\bibliographystyle{elsarticle-num}
\bibliography{mybib}

\end{document}